\documentclass[letterpaper, 10 pt, conference]{ieeeconf}  

\IEEEoverridecommandlockouts                              
\usepackage{graphics} 
\usepackage{epsfig} 
\title{\LARGE \bf
Tethered Flying Robot for Information Gathering System
}

\author{Tohru Ishii$^{1}$, Yasutake Takahashi$^{1}$, Yoichiro Maeda$^{2}$, and Takayuki Nakamura$^{3}$
\thanks{$^{1}$Tohru Ishii , Yasutake Takahashi and Yoichiro Maeda are with Department of Human and Artificial Intelligent Systems, Graduate School of Engineering, University of Fukui, 
		3-9-1, Bunkyo, Fukui, 910-8507, Japan
      {\tt\small \{tishii,yasutake\}@ir.his.u-fukui.ac.jp}}%
\thanks{$^{2}$Yoichiro Maeda is with Department of Robotics, Faculty of Engineering, Osaka Institute of Technology, 
		5-16-1, Omiya, Asahi-ku, Osaka, 535-8585, Japan
      {\tt\small maeda@bme.oit.ac.jp}}%
\thanks{$^{3}$Takayuki Nakamura is with Department of Computer and Communication Sciences, Faculty of Systems Engineering, Wakayama University, 
		Sakaetani 930, Wakayama 640-8510, Japan
      {\tt\small ntakayuk@sys.wakayama-u.ac.jp}
       }%
}

\begin{document}

\maketitle
\thispagestyle{empty}
\pagestyle{empty}

\begin{abstract}
Information from the sky is important for rescue activity in large-scale disaster or dangerous areas.
Observation system using a balloon or an airplane has been studied as an information gathering system from the sky. 
A balloon observation system needs helium gas and relatively long time to be ready.
An airplane observation system can be prepared in a short time and its mobility is good.
However, a long time flight is difficult because of limited amount of fuel. 

This paper proposes a kite-based observation system that complements activities of balloon and airplane observation systems by short preparation time and long time flight.
This research aims at construction of the autonomous flight information gathering system using a tethered flying unit that consists of the kite and the ground tether line control unit with a winding machine.
This paper reports development of the kite type tethered flying robot and an autonomous flying control system inspired by how to fly a kite by a human.
\end{abstract}

\section{Introduction}
Research and development of an information gathering system from the sky have done mainly for weather observation\cite{nishida1986,hirasawa1999,tnakao2005,ULDB2007}. 
Information gathering in large-scale disasters by manned aircraft has been becoming important because observation from the sky enables us to collect comprehensive information in one glance and it is usually hard to move around on the disaster affected ground.
In case that people have to keep away from the area, remote-controlled airplanes are often used to gather wide-range comprehensive information from the sky.
However, skilled remote operation is necessary to control the unmanned airplane safely.
Lack of remote operation skill often causes undesired accidents, therefore, fostering of skilled remote operators has been a vital problem nowadays.

Autonomous observation systems using a balloon\cite{zemi00,zemi01} or an airplane\cite{Karim:2005:EDI:1082473.1082799,jfujinaga2005,tsuzuki2008} have been studied as a solution of information gathering systems from the sky. 
The balloon system is noiseless and able to stay in the sky for a long time.
However, the helium gas reservation is necessary and it needs relatively long time and specialists of gas maintenance for the flight preparations.
On the other hand, an airplane system needs less time for flight preparations, but a long-term activity is difficult due to limitation of the fuel.

Although it is not an information gathering system, power generation systems using a kite, a balloon, and an airplane have been studied so far\cite{Engels2009a,Canale2007}. 
Yashwanth et al.\cite{mars} proposed to use a dynamo in a balloon, rotate a balloon by the power of a wind, and generate power. 
A balloon based power generation system tends to be a large scale to lift an electric dynamo in the sky. 
Power generation systems using a kite or an airplane\cite{Landrop2006,Lansdorp2007,5611288,Argatov20101052} control the kite or the airplane to have a certain trajectory so that they generate electric power by pulling the line connected to a power generator on the ground.
An airplane based power generation system needs to fly at high speed in order to pull the line connected to the electric dynamo efficiently and stay in the air.
It is useful to generate electric power during flying for a long-time self-contained observation, however, those system is not suitable for small-scale stationary observation and information gathering system because flying at high speed and high attitude is not so good for stationary observation and information gathering.

We propose a tethered flying robot based on a kite that flies with wind power as one of the natural power sources. 
It is supposed to be an information gathering system that compliments other ones based on a balloon or an airplane and has some advantage of short setup time and long-term observation.
This paper shows a prototype of the tethered flying robot we designed and built and its flight control system inspired by how a human flies a kite.
The kite-based flying robot is controlled with one line connected to a rotor on the ground by releasing or winding the line.
The robot has sensors that measure wind speeds, body posture and location in the air.
Preliminary experiments are conducted to confirm that the kite can take off from the ground autonomously and collect sensor data while the kite is flying.
Then, we built a simple flight control system that operates the line connected to the rotor on the ground based on the outputs of the wind sensors on the flight unit.
Real robot experiments are conducted to verify the performance of the flight controller.

\section{System Outline}
\begin{figure}[ht]
      \centering
      \includegraphics[width=0.3\textwidth]{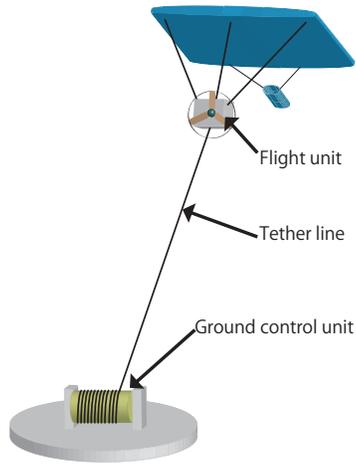}
      \caption{Concept of Tethered Flying Robot}
      \label{fig:system_outline}
\end{figure}

\begin{figure}[ht]
      \centering
      \includegraphics[width=0.3\textwidth]{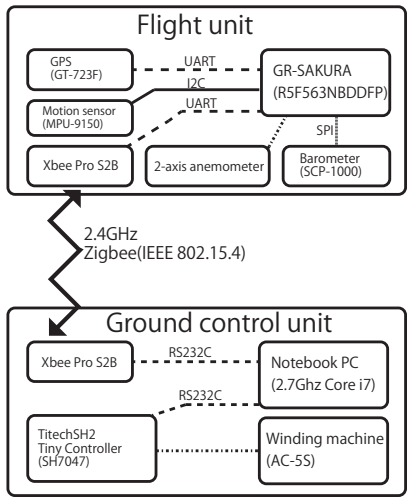}
      \caption{Schematic Representation of Tethered Flying Robot}
      \label{fig:system_sakura}
\end{figure}

Our tethered flying robot consists of a flight unit and a ground control unit. 
The concept of the flying robot and its schematic view are shown in Figs. \ref{fig:system_outline} and \ref{fig:system_sakura}, respectively.
The flight unit carries sensors and transmits the surrounding wind state and position and orientation of the flight unit itself to the ground wirelessly. 
The flight unit is lifted from the ground by a kite.
The ground control unit controls the line attached to the flying unit according to the data sent by the flight unit. 
A ZigBee module is used for wireless communication between the flight unit and the ground control unit. 


\subsection{Flight Unit}
\begin{figure}[ht]
      \centering
      \includegraphics[width=0.3\textwidth]{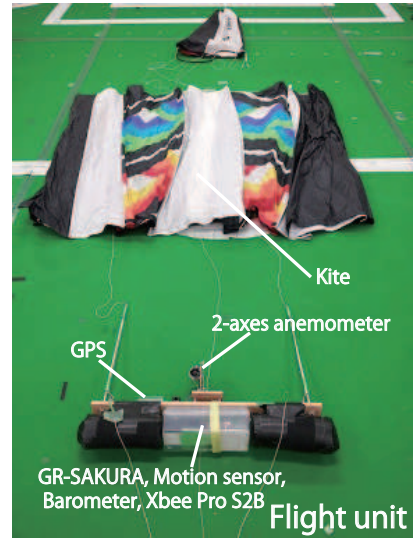}
      \caption{Flight Unit of Tethered Flying Robot}
      \label{fig:robot_main}
\end{figure}

The flight unit is shown in Fig. \ref{fig:robot_main}. 
The wingspan and the chord length of the kite are 3.2m and 1.5m, respectively.
The weight of the kite is about 700g.
The kite is able to carry about 1.5kg equipment in total.
The flight unit is equipped with anemometers to measure the wind speeds around the unit.
It is also equipped with a GPS for its own position estimation in the air, a accelerometer and a rate gyroscope for posture detection of the flight unit, and a ZigBee module (Xbee Pro S2B) for wireless data transmission to the ground unit.
It has an barometer so that the flight unit's altitude is calculated based on the atmospheric pressure difference from the ground. 
The weight of the flight unit is about 850g.

\begin{figure}[ht]
      \centering
      \includegraphics[width=0.3\textwidth]{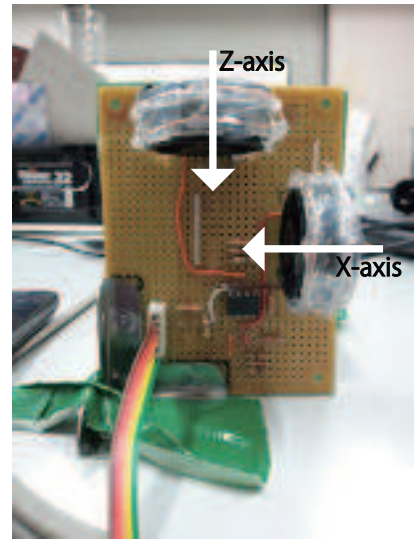}
      \caption{2-Axis Anemometer}
      \label{fig:anemometer}
\end{figure}
Fig.\ref{fig:anemometer} shows the two-axes anemometer that we designed and developed.
This anemometer uses impellers of commercially available portable anemometers and magnetic encoders. 
It can measure even weak wind because of very little resistance load.
Since the kite has only one line to be controlled, the flight unit is not able to control its posture by itself.
The anemometer is fixed on the flight unit and measures wind speed around the flight unit in orthogonal two directions.
Even when the flying unit inclines, the wind speed against the flight unit is calculated from the orthogonal two direction wind speeds and it is used for control.  

\subsection{Ground Control Unit}
\begin{figure}[ht]
      \centering
      \includegraphics[width=0.25\textwidth]{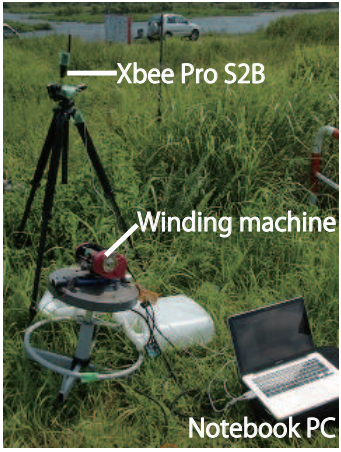}
      \caption{Ground Control Unit}
      \label{fig:robot_gcu}
\end{figure}

The ground control unit consists of a notebook computer and a line-winding machine.
The weight of the tether line is about 37.1g/100m.
Fig. \ref{fig:robot_gcu} shows the ground control unit which we developed.
The notebook computer receives the data from the flight unit and controls a line-winding machine through a microcomputer and a motor driver.
We adopt AC-5S manufactured by MIYAMAE Co.,Ltd which is originally an electric reel for sea fishing and modified it to the line-winding machine.
The line-winding machine has an electric clutch to adjust the brake strength of release of a line.
Moreover, it is possible to acquire information, including line length, number of rotations, etc., by the encoder in a line-winding machine. 
The length of the line set to the line-winding machine is about 300m.
The maximum winding-up power is 64kg and maximum take-up speed is 2.5m/s. 

\section{Takeoff Control}
\label{sec:TakeoffControl}
We developed two controllers for the tethered flying robot.
One is for a takeoff and the other is for staying in the air.
This section explains the takeoff controller and its experiment.

\begin{figure}[thpb]
      \centering
      \includegraphics[width=0.4\textwidth]{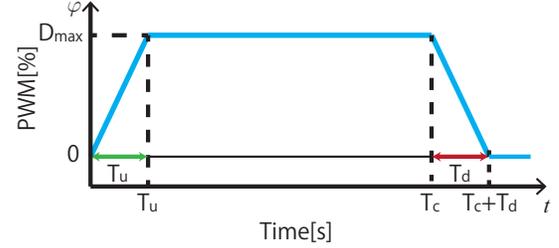}
      \caption{Winding Power (PWM duty ratio) Control during Takeoff}
      \label{fig:daikeictr}
\end{figure}

The experiment of the takeoff controller is conducted when the wind speed on the ground is zero so that we check the how high the flight unit can be rose with the controller. 
The line-winding machine winds the line according to a certain power profile.
The power profile is shown in Fig. \ref{fig:daikeictr}.
PWM duty ratio indicates the winding power in this case.
The line winding machine adds the power gradually until $T_{\rm u}$ s to the maximum power $D_{\max}$ \%.
It winds the line with the power $D_{\max}$ \% until $T_{\rm c}$ and reduce the power gradually.
This control equation is shown in (\ref{eq:control01}). 

\begin{eqnarray}
	\varphi (t) &=&
	\left\{ \begin{array}{l}
	D_{\max}t/T_{\rm u} \mbox{,} \hspace{10pt} \mbox{if } t \leq T_{u} \\
	D_{\max}  \mbox{,}\hspace{10pt} \mbox{if } (l_{\rm s}-l_{\rm p}<l) \mbox{and} (T_{\rm u}<t) \\
	D_{\max} - D_{\max}(t-T_{\rm c})/T_{\rm d} \mbox{,} \\ \hspace{40pt} \mbox{if } (t-T_{\rm c}) < T_{\rm d}\\
	0 \mbox{,} \hspace{32pt} \mbox{if }  (t-T_{\rm c}) \geq T_{\rm d}
	\end{array} \right.\label{eq:control01}
\end{eqnarray}
where $l_{\rm s}$,  $l$,  $l_{\rm p}$ are line length at the beginning, current, and the end, respectively.
$T_{\rm u}$ and $T_{\rm d}$ are duration of increasing and decreasing the PWM duty ratio at the beginning and the end, respectively.
$T_{\rm c} $ is the time in such a way that the line length $l$ reaches to the maximum length at the end $l_{\rm p}$.
$\varphi(t)$ and $D_{\max}$ are current and maximum PWM duty ratio, respectively.
$D_{\max}$ is set with 100 \% for this experiment. 

\begin{figure}[ht]
\begin{center}
      \includegraphics[width=0.45\textwidth]{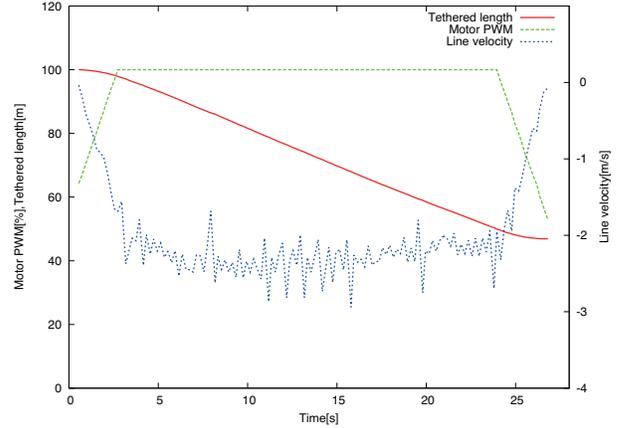}
      \caption{Winding Machine Log (Ground control unit)}
      \label{fig:daikeilog}
\end{center}
\end{figure}

\begin{figure}[thpb]
      \centering
      \includegraphics[width=0.45\textwidth]{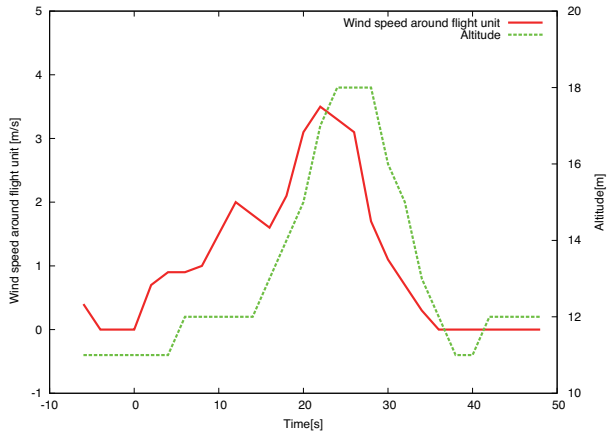}
      \caption{Flight Unit Log}
      \label{fig:daikeilog2}
\end{figure}

The experiment was conducted when wind speed was 0m/s. 
The kite cannot fly by itself at this wind speed.
We check how the flight unit takes off when the ground control unit winds the line as described above.
The line is wound up to 50 m from 100 m. 
The data log gathered by the ground control unit during the takeoff is shown in Fig. \ref{fig:daikeilog}. 
The wind speed around the flight unit and the altitude measured by the sensors on the flight unit are shown in Fig. \ref{fig:daikeilog2}. 
This record shows that the ground control unit operates to follow the takeoff controller.
The flight of about 7 m height has been checked according to the data. 
Fig. \ref{fig:daikeilog2} shows that altitude change of the flight unit has delay from change of wind speed.
The time lag between the altitude and the wind speed is about 4 seconds.

\section{Data Collection during Flight}
\begin{figure}[ht]
      \centering
      \includegraphics[width=0.45\textwidth]{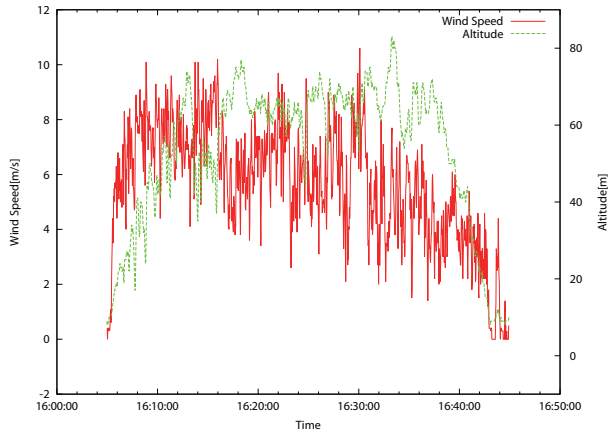}
      \caption{Flight log : Wind speed and altitude)}
      \label{fig:yobi_ws-alt}
\end{figure}

\begin{figure}[ht]
      \centering
      \includegraphics[width=0.4\textwidth]{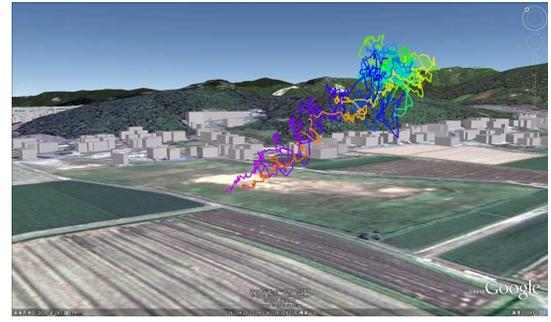}
      \caption{Flight log : GPS altitude\newline \copyright 2011 ZENRIN. Image \copyright 2011 GeoEye. GrayBuildings \copyright 2008 ZENRIN. \copyright 2011 Europa Technologies}
      \label{fig:log_gps_yobi}
\end{figure}

\begin{figure}[ht]
      \centering
      \includegraphics[width=0.45\textwidth]{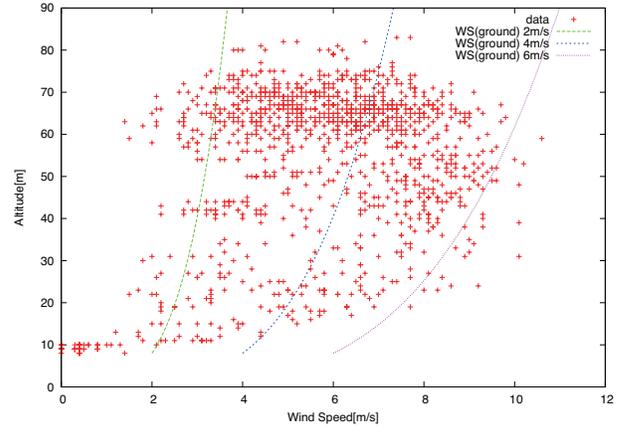}
      \caption{Relationship between wind speeds and altitude during flight}
      \label{fig:yobi_ws_alt}
\end{figure}

We check the ability of collecting data by sensors equipped on the flight unit while a human operator control the tether line.
The flight unit is controlled to stay at constant altitudes and acquires data at the each altitude.
In this experiment, wind speed, altitude and GPS data are collected.
Fig. \ref{fig:yobi_ws-alt} shows the wind speed and the altitude log.
Fig. \ref{fig:log_gps_yobi} shows a trail with the GPS and an altitude data during the flight.
Fig. \ref{fig:yobi_ws_alt} shows a relationship between wind speed and altitude during the flight.
A general relationship between wind speed and altitude according to wind speed at ground is also shown in Fig. \ref{fig:yobi_ws_alt}.
The wind speed in the sky varies from 2m/s to 10m/s according to Fig. \ref{fig:yobi_ws_alt}.
The kite itself can fly with wind speed at around 2.2m/s.
However, the flight unit needs at lest about 2.5m/s wind speed to stay in the air because of the weight of equipment.
Generally, the wind speed tends to be strong if altitude is high.
Therefore, the flight unit can be control to stay in the air for a long time stably as it flies at higher attitude in the sky.


\section{Flight Control based on Wind Speed and Experimental Results}
\begin{figure}[ht]
      \centering
      \includegraphics[width=0.35\textwidth]{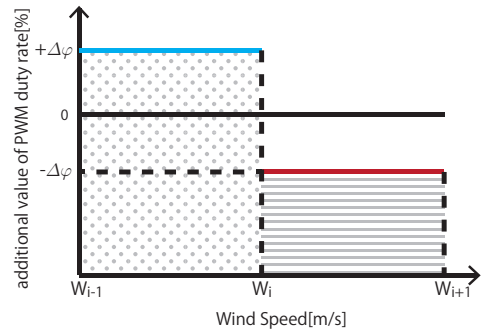}
      \caption{Basic idea of flight control}
      \label{fig:wsctr}
\end{figure}

The flight should be robust against the changes of wind speed during the flight.
The tether line should be controlled by the ground control unit accordingly otherwise the flight unit goes down when the wind becomes weak.
The proposed flight controller changes the tether line winding according to the wind speed around the flight unit measured by the anemometer on the flight unit.
The controller is developed and inspired by how a human flies a kite.
If the wind is strong, a line is released out then the flight unit is raised. 
If the wind is weak, a line is wound up to keep the attitude as much as possible. 
The basic idea of the flight control is shown in Fig. \ref{fig:wsctr}.
$W_{i}$ indicates threshold of wind speed around the flight unit.
$i$ is the index of thresholds.
It shows that the amount of increase and decrease of the PWM duty ratio is changed according to the wind speed. 
The PWM duty ratio becomes large to wind the line when a wind is weak as shown in the wind speed range from $W_{i-1}$ to $W_{i}$.
It decreases the PWM duty ratio to release the line in windy situation as shown in the wind speed range from $W_{i}$ to $W_{i+1}$.
This control equation is shown in (\ref{eq:control03}). 
The sampling time is 200ms for this experiment. 
\begin{equation}
	\varphi (t) = \varphi (t-1) + \Delta \varphi_{i-1} \;\;\;\;\; \mbox{if } W_{i-1} < W < W_{i} 
	\label{eq:control03}
\end{equation}
where $W$ and $\varphi (t)$ are the wind speed around the flight unit and PWM duty ratio, respectively.
$\Delta \varphi_{i-1}$ is additional value of PWM duty rate if the the wind speed $W$ is from $W_{i-1}$ to $W_{i}$ ($i=1,...,n$).

\begin{figure}[ht]
      \centering
      \includegraphics[width=0.45\textwidth]{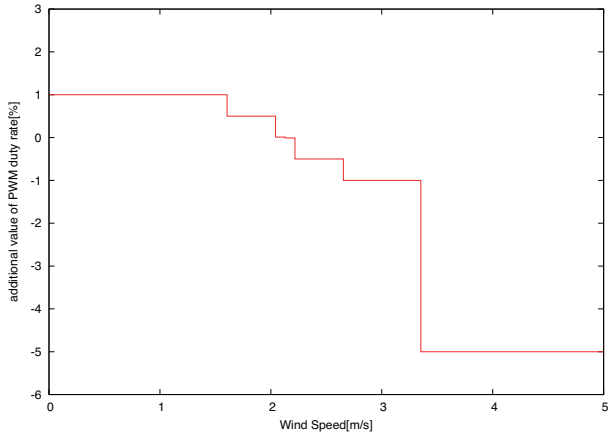}
      \caption{Flight Control Parameter}
      \label{fig:wsctr7}
\end{figure}

Experiments are carried out for showing the validity of the proposed flight control method.
The number of stages for changing the amount of increase and decrease of a duty ratio by wind speed around the flight unit $n$ is set to 7 and the additional value of PWM duty ratio $\Delta \varphi$ is set as shown in Fig. \ref{fig:wsctr7}. 
The thresholds of wind speeds $W_{i}$ and the number of stages $n$ are empirically determined. 

First, the tethered flying robot uses the takeoff control described in Section \ref{sec:TakeoffControl}.
Then, the experiment of the flight control based on the wind speed is started after it gradually releases the tether line to 100 m.

\begin{figure}[ht]
      \centering
      \includegraphics[width=0.45\textwidth]{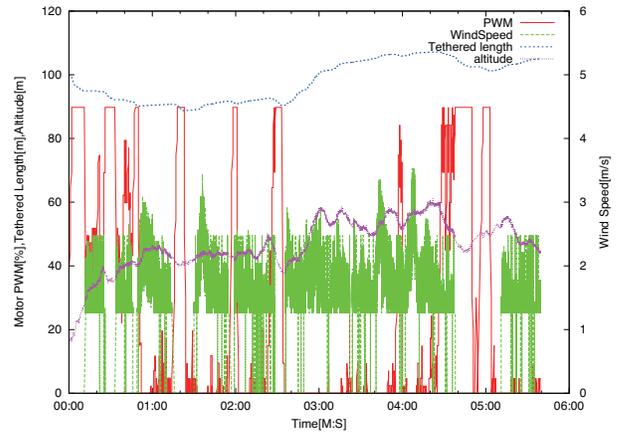}
      \caption{Flight log (Motor power, Wind speeds, Tethered line length, Altitude) during flight control based on wind speed}
      \label{fig:log}
\end{figure}

\begin{figure}[ht]
      \centering
      \includegraphics[width=0.45\textwidth]{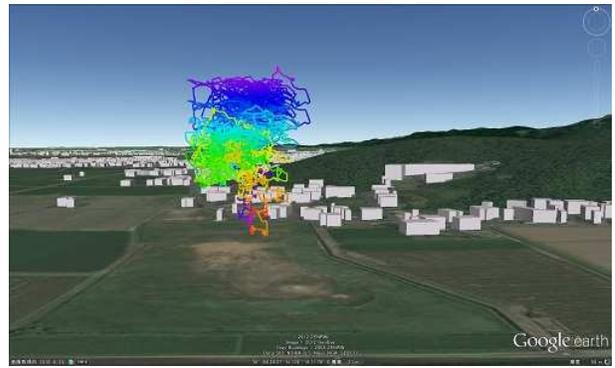}
      \caption{Flight log (GPS, altitude) during flight control based on wind speed \newline \copyright 2012 ZENRIN. Image \copyright GeoEye. GrayBuildings \copyright 2008 ZENRIN. Data SIO, NOAA, U.S. Navy, NGA, GEBCO}
      \label{fig:log_gps}
\end{figure}

\begin{figure}[thpb]
      \centering
      \includegraphics[width=0.45\textwidth]{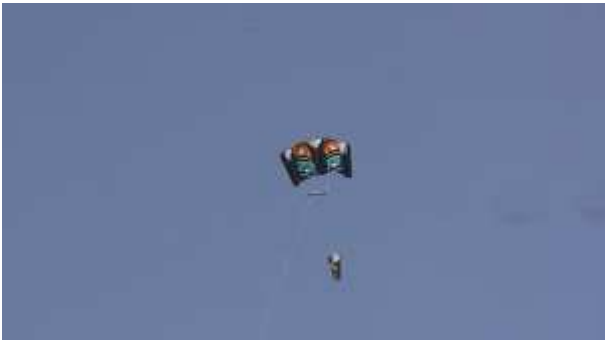}
      \caption{Picture of flight unit during flying}
      \label{fig:flying}
\end{figure}

Fig. \ref{fig:log} shows the flight log of PWM duty ratio, wind speed, line length, and the altitude.
Although the flight unit flies for about 60 minutes and collects data in this experiment, the data for the first about 6 minutes are shown here. 
Fig. \ref{fig:log_gps} shows the flight trajectory based on the GPS and the barometer outputs.
Fig. \ref{fig:flying} shows a picture of the flight unit while it is flying.

Since the wind is not blowing at the beginning, Fig. \ref{fig:log} shows the ground control unit winds the tether line to take the flight unit off from the ground.
Then, winding power is controlled according to the wind state around the flight unit measured by the anemometer on the unit.
The altitude of the flight unit gradually becomes higher.
After going up by about 20 m in 1 minute, it does not have strong wind enough to stay without winding the line.
Actually, it winds the line and it maintains the attitude.
The wind stops at about 4 minutes 30 seconds for a while and the ground control unit winds the line to avoid falling.
Consequently, the altitude of the flight unit is recovered gradually. 
The flight unit is flying on a swing right and left as shown in Fig. \ref{fig:log_gps}.
This is because the controller does not take a horizontal direction into consideration.


\section{Conclusions}
This research aims at realizing a flying observation system which complements other information gathering systems using a balloon or an air vehicle. 
We proposed the kite-based tethered flying robot with long-term activity capability.
It uses the wind as one of the natural energy.
In addition, it does not needs worries about fuel compared to an airplane-based system. 
The flight unit needs slight electric power.
Since a main controller is on the ground, it does not need to load a heavy battery on the flight unit.
It is able to supply electric power for the tethered flying robot itself if the flight unit carries dynamo aero generator and/or a solar panel. 
The proposed system has potential for a long-term flight.

This paper proposed a tethered flying robot and some experiment results.
The control of winding the line based on the wind speed effectively works and 60-minute-flight was realized with the simple controller inspired by how a human flies a kite.
For our future work, we would like to develop more sophisticated flight controller.






\section*{ACKNOWLEDGMENT}

This work was partially supported by JSPS KAKENHI Grant Number 24650118.

\bibliography{root}

\end{document}